\documentclass[aps,prl,twocolumn,showpacs,superscriptaddress]{revtex4}

\usepackage{graphicx}
\usepackage{dcolumn}
\usepackage{bm}
\usepackage{amssymb,amsmath}

\usepackage[usenames]{color}

\begin{document}

\newcommand{\dzero}     {D0}
\newcommand{\met}       {\mbox{$\not\!\!E_T$}}
\newcommand{\deta}      {\mbox{$\eta^{\rm det}$}}
\newcommand{\meta}      {\mbox{$\left|\eta\right|$}}
\newcommand{\mdeta}     {\mbox{$\left|\eta^{\rm det}\right|$}}
\newcommand{\rar}       {\rightarrow}
\newcommand{\rargap}    {\mbox{ $\rightarrow$ }}
\newcommand{\tbbar}     {\mbox{$tb$}}
\newcommand{\tqbbar}    {\mbox{$tqb$}}
\newcommand{\ttbar}     {\mbox{$t\bar{t}$}}
\newcommand{\bbbar}     {\mbox{$b\bar{b}$}}
\newcommand{\ccbar}     {\mbox{$c\bar{c}$}}
\newcommand{\qqbar}     {\mbox{$q\bar{q}$}}
\newcommand{\ppbar}     {\mbox{$p\bar{p}$}}
\newcommand{\comphep}   {\sc{c}\rm{omp}\sc{hep}}
\newcommand{\herwig}    {\sc{herwig}}
\newcommand{\pythia}    {\sc{pythia}}
\newcommand{\alpgen}    {\sc{alpgen}}
\newcommand{\singletop} {\rm{SingleTop}}
\newcommand{\reco}      {\sc{reco}}
\newcommand{\Mchiggs}   {\mbox{$M({\rm jet1,jet2},W)$}}
\newcommand{\coss}	{\mbox{\rm{cos}$\theta^{\star}$}}
\newcommand{\ljets} {$\ell+$jets}

\renewcommand{\arraystretch}{1.4}

\lefthyphenmin=6
\righthyphenmin=6

\hspace{5.2in}\mbox{Fermilab-Pub-12-096-E}

\title{Combination of searches for anomalous top quark couplings with 5.4~fb$^{\boldsymbol{-1}}$ of $\boldsymbol{p\bar{p}}$ collisions}
\affiliation{LAFEX, Centro Brasileiro de Pesquisas F\'{i}sicas, Rio de Janeiro, Brazil}
\affiliation{Universidade do Estado do Rio de Janeiro, Rio de Janeiro, Brazil}
\affiliation{Universidade Federal do ABC, Santo Andr\'e, Brazil}
\affiliation{University of Science and Technology of China, Hefei, People's Republic of China}
\affiliation{Universidad de los Andes, Bogot\'a, Colombia}
\affiliation{Charles University, Faculty of Mathematics and Physics, Center for Particle Physics, Prague, Czech Republic}
\affiliation{Czech Technical University in Prague, Prague, Czech Republic}
\affiliation{Center for Particle Physics, Institute of Physics, Academy of Sciences of the Czech Republic, Prague, Czech Republic}
\affiliation{Universidad San Francisco de Quito, Quito, Ecuador}
\affiliation{LPC, Universit\'e Blaise Pascal, CNRS/IN2P3, Clermont, France}
\affiliation{LPSC, Universit\'e Joseph Fourier Grenoble 1, CNRS/IN2P3, Institut National Polytechnique de Grenoble, Grenoble, France}
\affiliation{CPPM, Aix-Marseille Universit\'e, CNRS/IN2P3, Marseille, France}
\affiliation{LAL, Universit\'e Paris-Sud, CNRS/IN2P3, Orsay, France}
\affiliation{LPNHE, Universit\'es Paris VI and VII, CNRS/IN2P3, Paris, France}
\affiliation{CEA, Irfu, SPP, Saclay, France}
\affiliation{IPHC, Universit\'e de Strasbourg, CNRS/IN2P3, Strasbourg, France}
\affiliation{IPNL, Universit\'e Lyon 1, CNRS/IN2P3, Villeurbanne, France and Universit\'e de Lyon, Lyon, France}
\affiliation{III. Physikalisches Institut A, RWTH Aachen University, Aachen, Germany}
\affiliation{Physikalisches Institut, Universit\"at Freiburg, Freiburg, Germany}
\affiliation{II. Physikalisches Institut, Georg-August-Universit\"at G\"ottingen, G\"ottingen, Germany}
\affiliation{Institut f\"ur Physik, Universit\"at Mainz, Mainz, Germany}
\affiliation{Ludwig-Maximilians-Universit\"at M\"unchen, M\"unchen, Germany}
\affiliation{Fachbereich Physik, Bergische Universit\"at Wuppertal, Wuppertal, Germany}
\affiliation{Panjab University, Chandigarh, India}
\affiliation{Delhi University, Delhi, India}
\affiliation{Tata Institute of Fundamental Research, Mumbai, India}
\affiliation{University College Dublin, Dublin, Ireland}
\affiliation{Korea Detector Laboratory, Korea University, Seoul, Korea}
\affiliation{CINVESTAV, Mexico City, Mexico}
\affiliation{Nikhef, Science Park, Amsterdam, the Netherlands}
\affiliation{Radboud University Nijmegen, Nijmegen, the Netherlands}
\affiliation{Joint Institute for Nuclear Research, Dubna, Russia}
\affiliation{Institute for Theoretical and Experimental Physics, Moscow, Russia}
\affiliation{Moscow State University, Moscow, Russia}
\affiliation{Institute for High Energy Physics, Protvino, Russia}
\affiliation{Petersburg Nuclear Physics Institute, St. Petersburg, Russia}
\affiliation{Instituci\'{o} Catalana de Recerca i Estudis Avan\c{c}ats (ICREA) and Institut de F\'{i}sica d'Altes Energies (IFAE), Barcelona, Spain}
\affiliation{Uppsala University, Uppsala, Sweden}
\affiliation{Lancaster University, Lancaster LA1 4YB, United Kingdom}
\affiliation{Imperial College London, London SW7 2AZ, United Kingdom}
\affiliation{The University of Manchester, Manchester M13 9PL, United Kingdom}
\affiliation{University of Arizona, Tucson, Arizona 85721, USA}
\affiliation{University of California Riverside, Riverside, California 92521, USA}
\affiliation{Florida State University, Tallahassee, Florida 32306, USA}
\affiliation{Fermi National Accelerator Laboratory, Batavia, Illinois 60510, USA}
\affiliation{University of Illinois at Chicago, Chicago, Illinois 60607, USA}
\affiliation{Northern Illinois University, DeKalb, Illinois 60115, USA}
\affiliation{Northwestern University, Evanston, Illinois 60208, USA}
\affiliation{Indiana University, Bloomington, Indiana 47405, USA}
\affiliation{Purdue University Calumet, Hammond, Indiana 46323, USA}
\affiliation{University of Notre Dame, Notre Dame, Indiana 46556, USA}
\affiliation{Iowa State University, Ames, Iowa 50011, USA}
\affiliation{University of Kansas, Lawrence, Kansas 66045, USA}
\affiliation{Kansas State University, Manhattan, Kansas 66506, USA}
\affiliation{Louisiana Tech University, Ruston, Louisiana 71272, USA}
\affiliation{Boston University, Boston, Massachusetts 02215, USA}
\affiliation{Northeastern University, Boston, Massachusetts 02115, USA}
\affiliation{University of Michigan, Ann Arbor, Michigan 48109, USA}
\affiliation{Michigan State University, East Lansing, Michigan 48824, USA}
\affiliation{University of Mississippi, University, Mississippi 38677, USA}
\affiliation{University of Nebraska, Lincoln, Nebraska 68588, USA}
\affiliation{Rutgers University, Piscataway, New Jersey 08855, USA}
\affiliation{Princeton University, Princeton, New Jersey 08544, USA}
\affiliation{State University of New York, Buffalo, New York 14260, USA}
\affiliation{Columbia University, New York, New York 10027, USA}
\affiliation{University of Rochester, Rochester, New York 14627, USA}
\affiliation{State University of New York, Stony Brook, New York 11794, USA}
\affiliation{Brookhaven National Laboratory, Upton, New York 11973, USA}
\affiliation{Langston University, Langston, Oklahoma 73050, USA}
\affiliation{University of Oklahoma, Norman, Oklahoma 73019, USA}
\affiliation{Oklahoma State University, Stillwater, Oklahoma 74078, USA}
\affiliation{Brown University, Providence, Rhode Island 02912, USA}
\affiliation{University of Texas, Arlington, Texas 76019, USA}
\affiliation{Southern Methodist University, Dallas, Texas 75275, USA}
\affiliation{Rice University, Houston, Texas 77005, USA}
\affiliation{University of Virginia, Charlottesville, Virginia 22901, USA}
\affiliation{University of Washington, Seattle, Washington 98195, USA}
\author{V.M.~Abazov} \affiliation{Joint Institute for Nuclear Research, Dubna, Russia}
\author{B.~Abbott} \affiliation{University of Oklahoma, Norman, Oklahoma 73019, USA}
\author{B.S.~Acharya} \affiliation{Tata Institute of Fundamental Research, Mumbai, India}
\author{M.~Adams} \affiliation{University of Illinois at Chicago, Chicago, Illinois 60607, USA}
\author{T.~Adams} \affiliation{Florida State University, Tallahassee, Florida 32306, USA}
\author{G.D.~Alexeev} \affiliation{Joint Institute for Nuclear Research, Dubna, Russia}
\author{G.~Alkhazov} \affiliation{Petersburg Nuclear Physics Institute, St. Petersburg, Russia}
\author{A.~Alton$^{a}$} \affiliation{University of Michigan, Ann Arbor, Michigan 48109, USA}
\author{G.~Alverson} \affiliation{Northeastern University, Boston, Massachusetts 02115, USA}
\author{M.~Aoki} \affiliation{Fermi National Accelerator Laboratory, Batavia, Illinois 60510, USA}
% ...
%
%
\collaboration{The D0 Collaboration\footnote{with visitors from
$^{a}$Augustana College, Sioux Falls, SD, USA,
$^{b}$The University of Liverpool, Liverpool, UK,
$^{c}$UPIITA-IPN, Mexico City, Mexico,
$^{d}$DESY, Hamburg, Germany,
,
$^{e}$SLAC, Menlo Park, CA, USA,
$^{f}$University College London, London, UK,
$^{g}$Centro de Investigacion en Computacion - IPN, Mexico City, Mexico,
$^{h}$ECFM, Universidad Autonoma de Sinaloa, Culiac\'an, Mexico
and
$^{i}$Universidade Estadual Paulista, S\~ao Paulo, Brazil.
$^{\ddag}$Deceased.
}} \noaffiliation
\vskip 0.25cm

\date{April 10, 2012}

\begin{abstract}
We present measurements of the $tWb$ coupling form factors using information from electroweak
single top quark production and from the helicity of $W$~bosons from top quark decays in 
{\ttbar} events. We set upper limits on anomalous $tWb$ coupling form factors using data 
collected with the D0 detector at the Tevatron $p\bar{p}$ collider corresponding
to an integrated luminosity of 5.4 fb$^{-1}$.


\end{abstract}
\maketitle

\vspace{-0.1in}
The top quark is being studied in unprecedented detail with the large data samples 
from Run II of the Fermilab Tevatron collider. Since the top quark is by far the most massive 
known fermion, with a coupling to the Higgs field of order unity, these studies may shed light 
on the mechanism of electroweak symmetry breaking and provide hints of new physics.
Within the standard model (SM), the top quark coupling to the 
bottom quark and the $W$~boson ($tWb$) has the $V-A$ form of a left-handed vector interaction.
We consider a more general form for the $tWb$ coupling to allow for departures 
from the SM~\cite{TT_window}. We look for physics beyond the SM in the form
of right-handed vector couplings or left- or right-handed tensor couplings, described
by the effective Lagrangian including operators up to dimension five~\cite{cpyuan_0503040v3}:
\begin{eqnarray}
\mathcal{L}&=& \frac{g}{\sqrt{2}}\bar{b} \gamma^\mu V_{tb}
\nonumber (f^L_V P_L + f^R_V P_R) t W_{\mu}^{-}\\
& &-\frac{g}{\sqrt{2}} \bar{b} \frac{i\sigma^{\mu\nu} q_{\nu}V_{tb}}{M_W} 
(f^L_T P_L + f^R_T P_R) t W_{\mu}^{-} + h.c. \, ,
\label{coupling}
\end{eqnarray}
 where  $M_W$ is the mass of the $W$~boson, $q$ is its four-momentum, $V_{tb}$ is the 
Cabibbo-Kobayashi-Maskawa matrix element~\cite{Cabibbo:1963yz}, and $P_{L}=(1 - \gamma_5)/2$ 
($P_{R}=(1 + \gamma_5)/2$) is the left-handed (right-handed) projection operator. In the SM,
the left-handed vector coupling form factor is $f^L_V=1$, the right-handed vector coupling 
form factor is $f^R_V=0$, and the tensor coupling form factors are $f^L_T=f^R_T=0$. 
We assume real coupling form factors, implying $CP$ conservation, 
and a spin-$\frac{1}{2}$ top quark which decays predominantly to $Wb$.

An alternative parameterization of anomalous couplings through effective operators has
been proposed recently~\cite{Willenbrock,Aguilar}. The anomalous coupling limits presented
in this letter can be translated into the operator parameterization~\cite{Aguilar}:
\begin{eqnarray}
\label{eq:operator}
|f^L_V| &=& 1 + |C_{\phi q}^{(3,3+3)}| \frac{v^2}{V_{tb} \Lambda^2} \; ,\nonumber \\
|f^R_V| &=& \frac{1}{2} |C_{\phi \phi}^{33}| \frac{v^2}{V_{tb} \Lambda^2} \; ,\nonumber \\
|f^L_T| &=& \sqrt{2} |C_{dW}^{33}| \frac{v^2}{V_{tb} \Lambda^2} \; ,\nonumber \\
|f^R_T| &=& \sqrt{2} |C_{uW}^{33}| \frac{v^2}{V_{tb} \Lambda^2} \; , 
\end{eqnarray}
where $\Lambda$ is the scale of the new physics and $v=246$~GeV is the scale
of electroweak symmetry breaking. $C_{\phi q}^{(3,3+3)}$,
$C_{\phi \phi}^{33}$, $C_{dW}^{33}$ and $C_{uW}^{33}$ are constants
for dimension-six gauge-invariant effective operators for third generation quarks,
involving the Higgs field ($\phi$),
the $W$~boson, up-type ($u$) and down-type ($d$) quarks. 
The constants $C$ are assumed to be real.

Indirect constraints on the magnitude of the right-handed vector coupling and tensor 
couplings exist from measurements of the $b \rightarrow s\gamma$ branching 
fraction~\cite{bsgamma}. General unitarity considerations require the anomalous 
tensor couplings to be less than 0.5~\cite{Gounaris:1995ce}.
While the  $b \rightarrow s\gamma$ limits are tighter than the direct limits presented 
in this Letter, 
they include assumptions that are not required here, in particular that there is
no new physics affecting the $b$~quark other than anomalous $tWb$ couplings. 
Direct constraints on anomalous $tWb$ couplings have been obtained from previous 
D0 analyses~\cite{Abazov:2011pm,Abazov:2009ky}
and from an analysis of LHC results~\cite{AguilarSaavedra:2011ct}.

This Letter describes a combination of recent $W$~boson helicity~\cite{Abazov:2010jn} and
single top quark~\cite{Abazov:2011pm} measurements, using the same procedure as in a previous 
combination of $W$~boson helicity with single top quark information in D0 
data~\cite{Abazov:2009ky}.
Deviations from the SM expectation in the coupling form factors manifest themselves in two 
distinct ways that are observable at D0: (i) by altering the fractions
of $W$~bosons from top quark decays produced in each of the three possible helicity 
states, and (ii) by changing the rate and kinematic 
distributions of electroweak single top quark production. We translate $W$~boson helicity 
fractions~\cite{Abazov:2010jn} into form factors using the general framework given in 
Ref.~\cite{Chen:2005vr}. By combining these with the
single top quark anomalous couplings analysis~\cite{Abazov:2011pm}, we obtain
posterior probability density distributions for the anomalous coupling form factors.
Three separate scenarios are investigated using the same dataset, for $f^R_V$, $f^L_T$, and
$f^R_T$. In each scenario we investigate the anomalous coupling form
factor and the SM coupling form factor $f^L_V$ simultaneously and set the other two anomalous
coupling form factors to zero. 
We form a two-dimensional posterior density as a function of two coupling
form factors and then marginalize over the SM coupling to obtain a 95\% C.L. limit on the 
anomalous coupling.

This analysis is based on data collected with the D0 
detector~\cite{NIM,Angstadt:2009ie,Ahmed:2010fx,Abolins:trigger} corresponding to an 
integrated luminosity of 5.4~fb$^{-1}$. For the $W$~boson helicity analysis, {\ttbar} 
events are selected in both the lepton plus jets 
($t\bar{t}\rightarrow W^+W^-b\bar{b}\rightarrow \ell\nu q\bar{q}^{\prime}b\bar{b}$, requiring
a lepton, missing transverse energy and at least four jets) and dilepton 
($t\bar{t}\rightarrow W^+W^-b\bar{b}\rightarrow \ell\nu\ell^{\prime}\nu^{\prime}b\bar{b}$,
requiring two leptons, missing transverse energy and at least two jets)
channels~\cite{Abazov:2010jn}.

We use the {\alpgen} leading-order Monte Carlo (MC) event generator~\cite{alpgen}, 
interfaced to {\sc pythia}~\cite{pythia}, to model {\ttbar} events as well as
$W$+jets and $Z$+jets background events. We generate {\ttbar} events with both 
SM $V-A$ and $V+A$ couplings, and reweight these to model any given $W$~boson 
helicity state. 
We use the CTEQ6L1 parton distribution functions~\cite{cteq} and set the top quark mass to 
172.5~GeV, consistent with the world average top mass~\cite{TEWG:2011wr}.
The response of the D0 detector is simulated using {\sc geant}~\cite{geant}. 
The presence of additional {\ppbar} interactions is modeled by overlaying the simulation with 
data events, selected from random 
beam crossings matching the instantaneous luminosity profile in the data.
The background 
from multijet production, where a jet is misidentified as an isolated 
electron or muon, is modeled with data events containing lepton candidates that
pass all of the lepton identification requirements except one, but otherwise resemble the 
signal events. We use MC to model the smaller background from dibosons.
The SM single top quark background is modeled using the {\sc comphep} MC event 
generator~\cite{singletop-mcgen} normalized to theory 
predictions~\cite{singletop-xsec-kidonakis}.
In the $W$~boson helicity analysis, the possible presence of anomalous couplings does not 
significantly modify the small background from single top quark production.
A multivariate likelihood discriminant that uses both 
kinematic and $b$~quark lifetime information distinguishes {\ttbar} events from background,
separately in the lepton plus jets and dilepton channels. A requirement on the likelihood
selects 1431 lepton plus jet events and 319 dilepton events with expected backgrounds
of $404 \pm 32$ and $69 \pm 10$ events, respectively, where the uncertainty includes
both statistical and background modeling components.

\begin{figure}[!h!btp]
\includegraphics[width=0.46\textwidth]{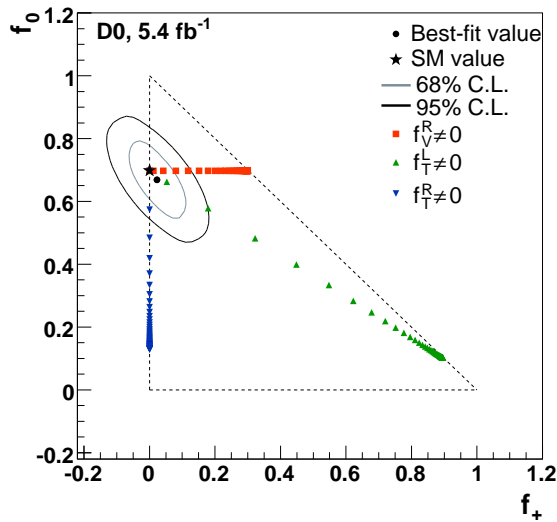}
\vspace{-0.2in}
\caption{(color online) Likelihood contours at the 68\% C.L. and the 95\% C.L.
as a function of $W$~boson helicity fractions. 
Statistical uncertainties and systematic 
uncertainties that are uncorrelated with the single top quark measurement are included.
The squares, triangles and upside-down triangles
show $f_V^R$, $f_T^L$ and $f_T^R$ varying in fifty equal-size steps such that their ratio 
to $f_V^L$ goes from zero to ten-to-one.
The dashed triangle denotes the physically allowed region.
 }
\label{fig:whel_map}
\end{figure}

We determine the fractions of $W$ bosons with left-handed, longitudinal, and right-handed 
helicity ($f_-$, $f_0$, and $f_+$, respectively). 
The SM predicts  $f_-=30$\%, $f_0=70$\%, and $f_+\approx {\cal O}(10^{-4})$~\cite{Czarnecki}. 
The fractions are measured in a fit to the distribution of the angle $\theta^*$,
where $\theta^*$ is the angle between the direction 
opposite to the top quark and the direction of the down-type fermion (charged lepton or 
down-type quark) from the decay of the $W$~boson, both in the $W$~boson rest frame.
A binned maximum likelihood fit compares the $\cos\theta^*$ distribution of the selected 
events to expectations from each $W$~boson helicity state and the background.
In the lepton plus jets channel, each possible assignment of the four leading jets in 
the event is considered to reconstruct the two top quarks in the event, based on the 
$\chi^2$ of a kinematic fit and the compatibility between the assigned jet flavor and 
$b$~quark lifetime information. For the $W$~boson that decays hadronically, we do not attempt to 
determine which of the daughter jets corresponds to the up-type quark. Rather we select one 
jet at random. Since this introduces a sign ambiguity, we can only distinguish the longitudinal
helicity from the other two states and can no longer distinguish left-handed and 
right-handed helicity states. 
In the dilepton channel, we determine the momenta of the two neutrinos using an
algebraic solution. Since the system is kinematically underconstrained, we assume a
value for the top quark mass of 172.5~GeV to perform the kinematic reconstruction. 
We vary both the longitudinal and right-handed helicity fractions $f_0$ and $f_+$ in the 
fit and find the relative likelihood of any set of helicity fractions being consistent 
with the data.  The result is presented in Fig.~\ref{fig:whel_map}, which also demonstrates 
how non-SM values for the coupling form factors could alter the $W$~boson helicity fractions.  

\begin{figure*}[!h!btp]
\includegraphics[width=0.32\textwidth]{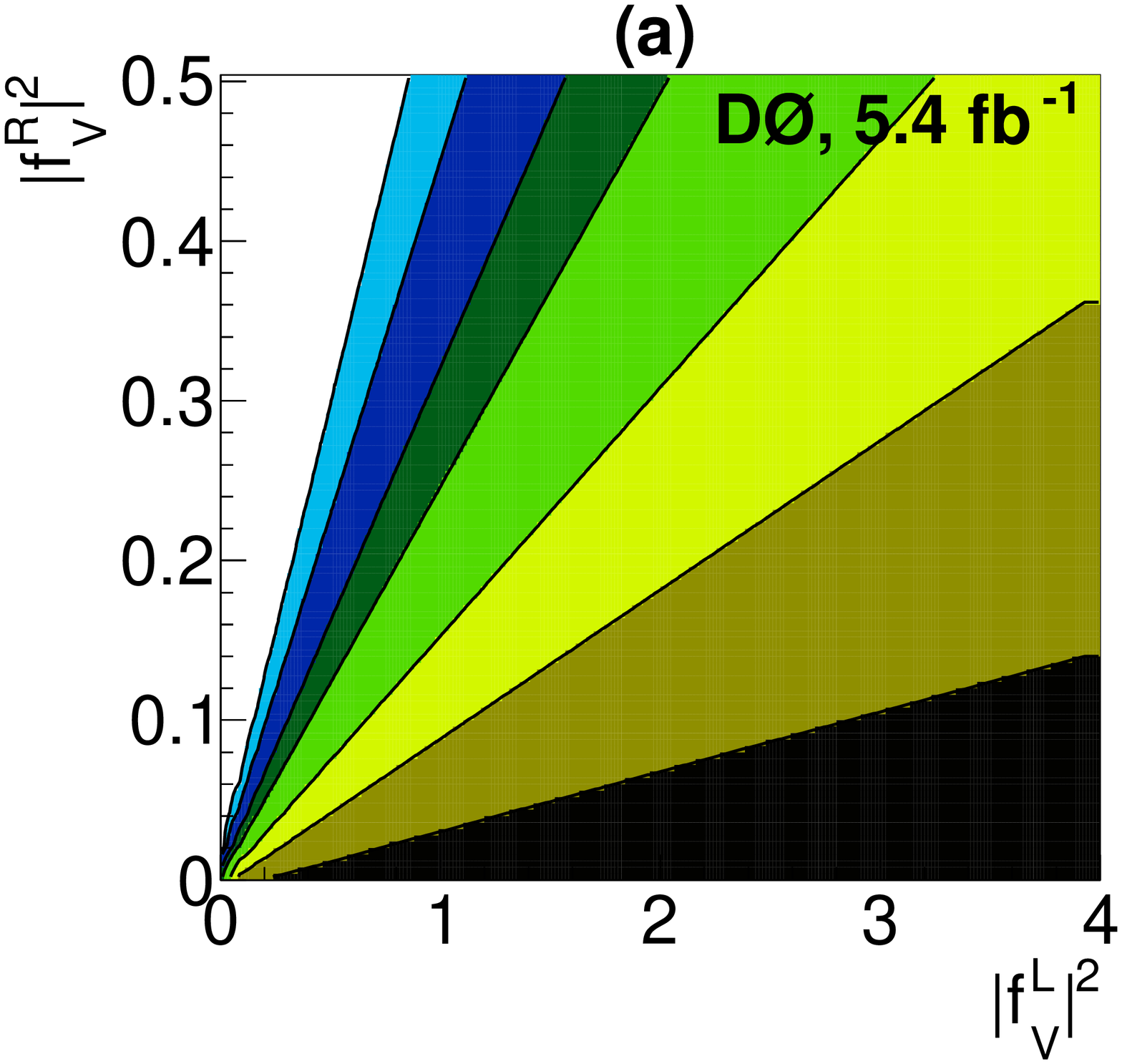}
\includegraphics[width=0.32\textwidth]{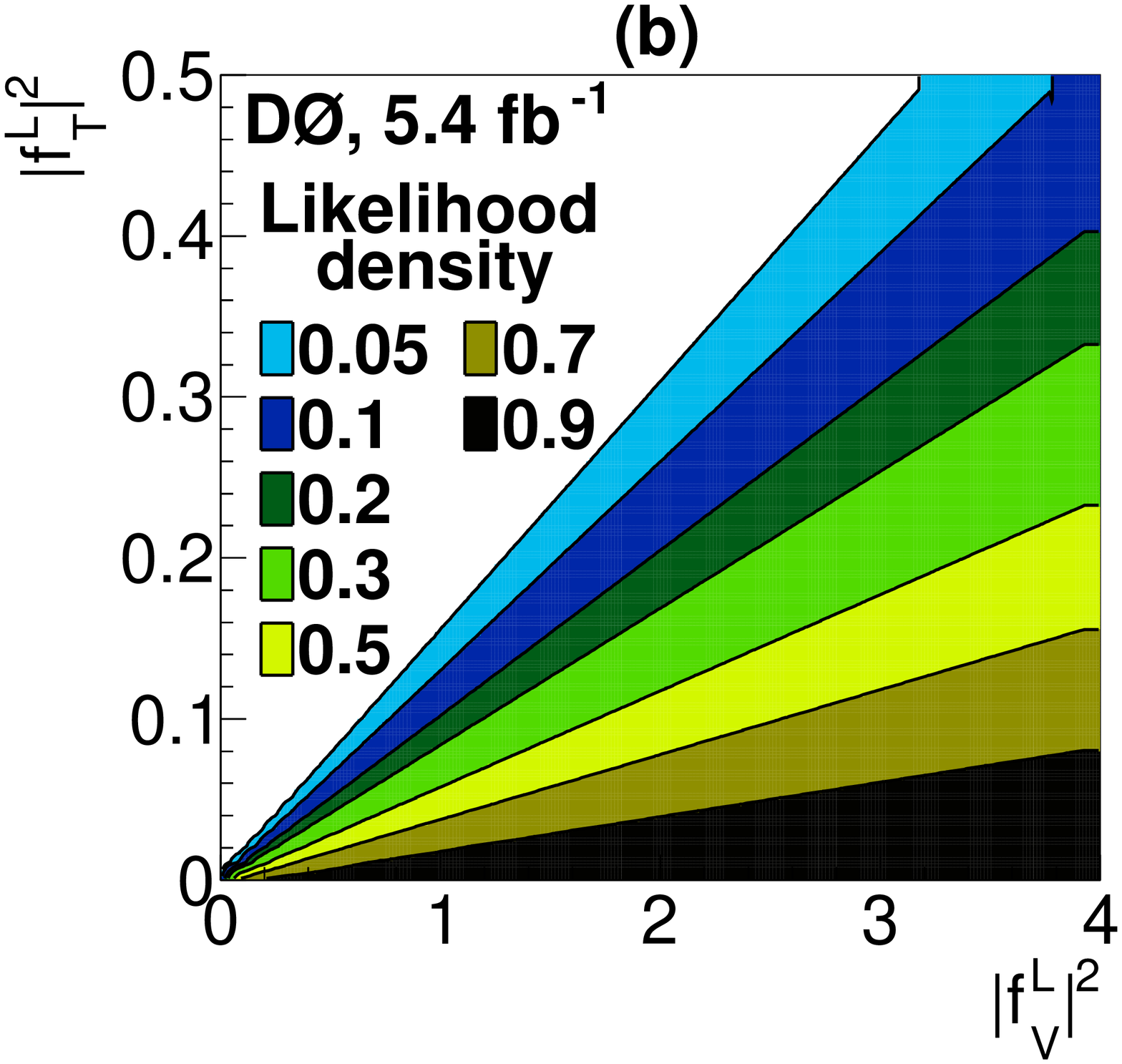}
\includegraphics[width=0.32\textwidth]{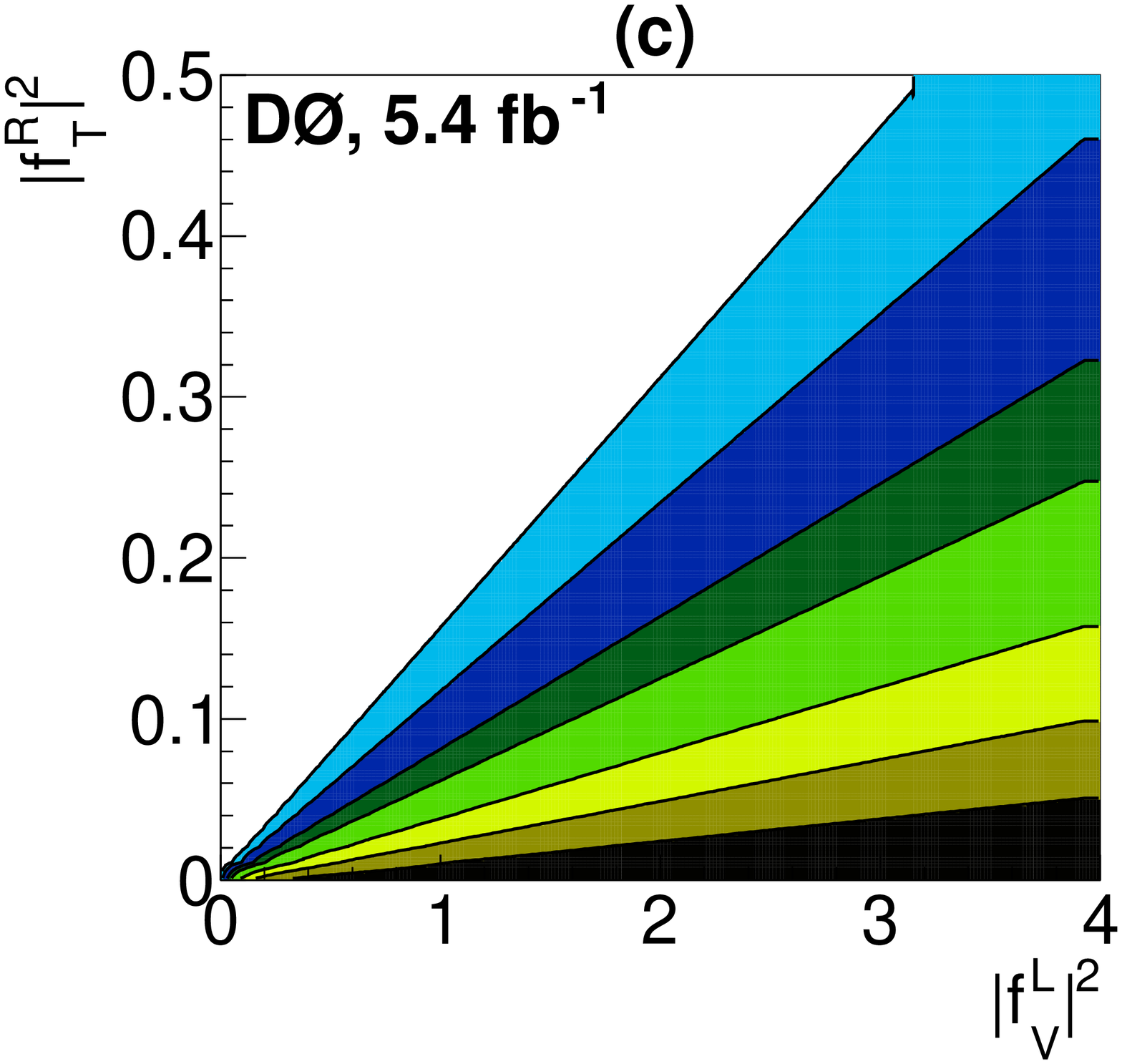}
\vspace{-0.1in}
\caption{(color online) Likelihood density as a function of $tWb$ coupling form factors, 
for (a) right-vector vs. left-vector couplings,
(b) left-tensor vs. left-vector couplings, and
(c) right-tensor vs. left-vector couplings, using information from the 
$W$~boson helicity measurement only. All systematic uncertainties are included. Each color 
corresponds to a contour of equal likelihood density. }
\label{fig:whel_prior}
\end{figure*}

The result is interpreted in terms of the coupling form factors in Fig.~\ref{fig:whel_prior}, 
which
shows that the $W$~boson helicity measurement only constrains ratios of the coupling form 
factors and not their magnitude. These distributions provide one of the inputs to the
combined constraint on the coupling form factors.


The other input to the form factor constraint comes from the search for anomalous
$tWb$ couplings in the single top quark final state. Both $t$-channel (the exchange of a 
$W$~boson between a light quark and a heavy quark) and $s$-channel (the production and
decay of a virtual $W$~boson) modes contribute to
single top quark production at the Tevatron. Single top quark production was observed by the 
CDF and D0 collaborations~\cite{stop-obs-2009-d0,stop-obs-2009-cdf}, and the $t$-channel
mode was also isolated by the D0 collaboration~\cite{t-channel-new}.

Both the single top quark production cross section and kinematic distributions are modified
by anomalous couplings. The single top quark cross section may also differ from the 
SM prediction because $|V_{tb}|<1$, but that is not considered here.
We assume that single top quark production proceeds 
exclusively through the $tWb$ vertex and not through the exchange of a new particle.
We also assume that $|V_{td}|^2+|V_{ts}|^2 \ll |V_{tb}|^2$, i.e., top quark production
and decay through light quarks is negligible.

The single top quark anomalous couplings analysis selects events in which the top quark 
decays to a $W$~boson and a $b$~quark, followed by the decay of the $W$~boson to an electron
or muon, and a neutrino. The final state contains two or three jets, one from the top quark 
decay, one produced together with the top quark, and possibly a third jet from initial-state or
final-state gluon radiation. The event selection is identical to that in the 
anomalous coupling single top quark analysis~\cite{Abazov:2011pm} and the SM single top quark
analysis~\cite{Abazov:2011pt}, except that events with four jets are removed from the sample
to avoid overlap with the $W$~boson helicity analysis. One or two of the jets are required to 
be $b$-tagged, i.e., identified as originating from $B$~hadrons~\cite{D0btag}. 
To increase the search sensitivity, the data are divided into four independent analysis 
channels based on jet multiplicity (2 or 3), and number of $b$-tagged jets (1 or 2). 

We use Bayesian neural networks (BNN)~\cite{bayesianNNs} to  discriminate 
between the single top quark anomalous coupling signal and the backgrounds. For each of the 
three coupling scenarios, the signal in the BNN training consists of only that 
particular anomalous single top quark couplings sample while the background in the training 
consists of all SM backgrounds plus SM single top quark events.
The main background contributions to the single top quark analysis are those from $W$+jets, 
{\ttbar} and multijet
production. The background modeling and normalization procedures are the same as in the
$W$~boson helicity analysis. The {\ttbar} contribution to the background is small and is
modeled by simulated SM {\ttbar} events and normalized to the theoretical cross
section~\cite{ttbar-xsec}. The effect of anomalous couplings on the {\ttbar} background is
negligible. We model the single top quark signal using the {\sc comphep} MC event
generator~\cite{singletop-mcgen} where anomalous $tWb$ couplings are considered in both
the production and decay of the top quark.

We use the four-vectors of the reconstructed final state particles in the BNN training
(transverse momentum $p_T$, pseudorapidity $\eta$, angle $\Delta \phi$ with respect to the
lepton, and the mass of each jet), i.e., twelve variables for events with two jets and 
sixteen variables for events with three jets.
We add four angular variables that are particularly sensitive to the anomalous couplings.
These are cosines of angles between various final state objects in the top quark rest frame.

\begin{figure*}[!h!btp]
\includegraphics[width=0.32\textwidth]{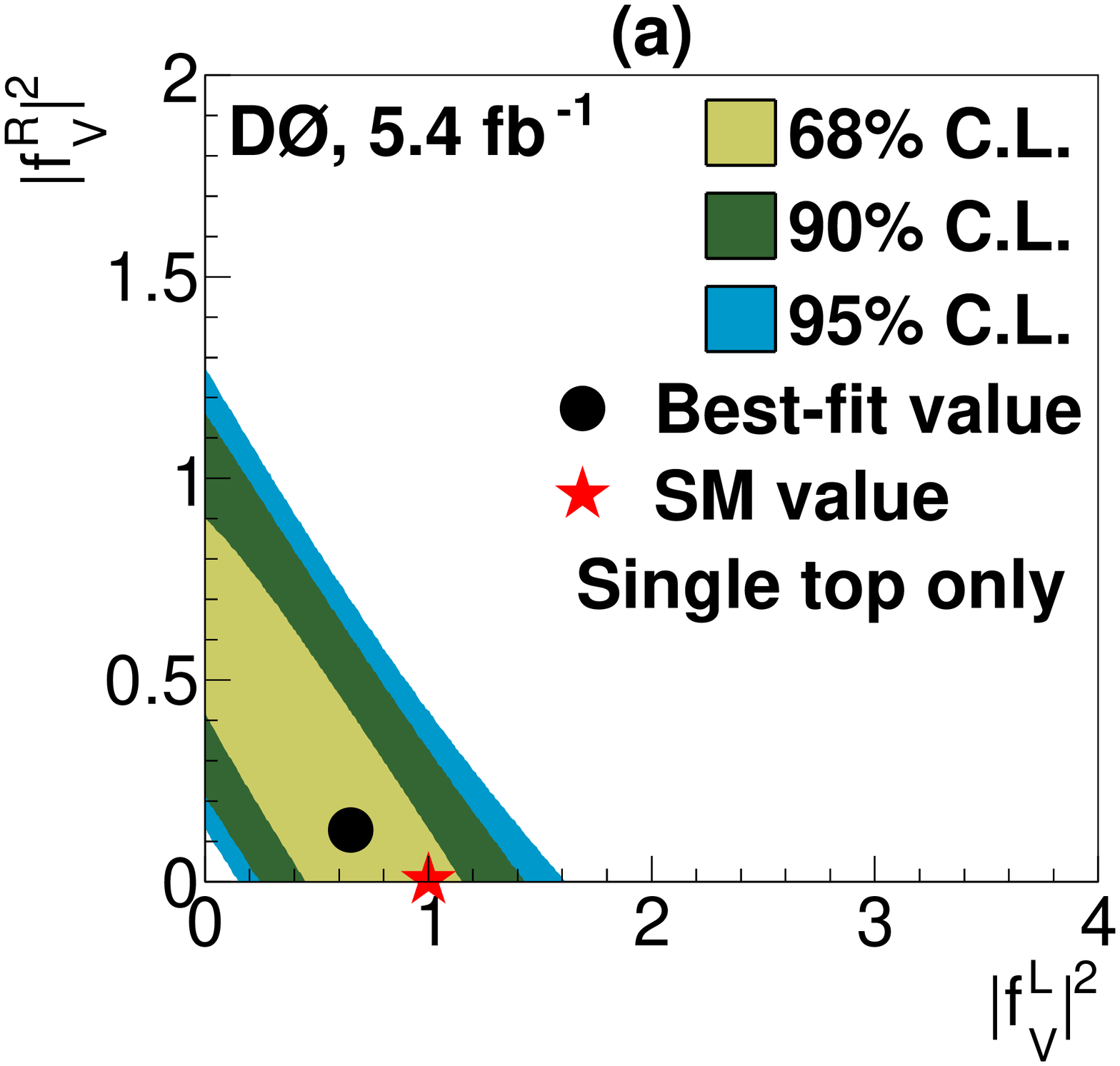}
\includegraphics[width=0.32\textwidth]{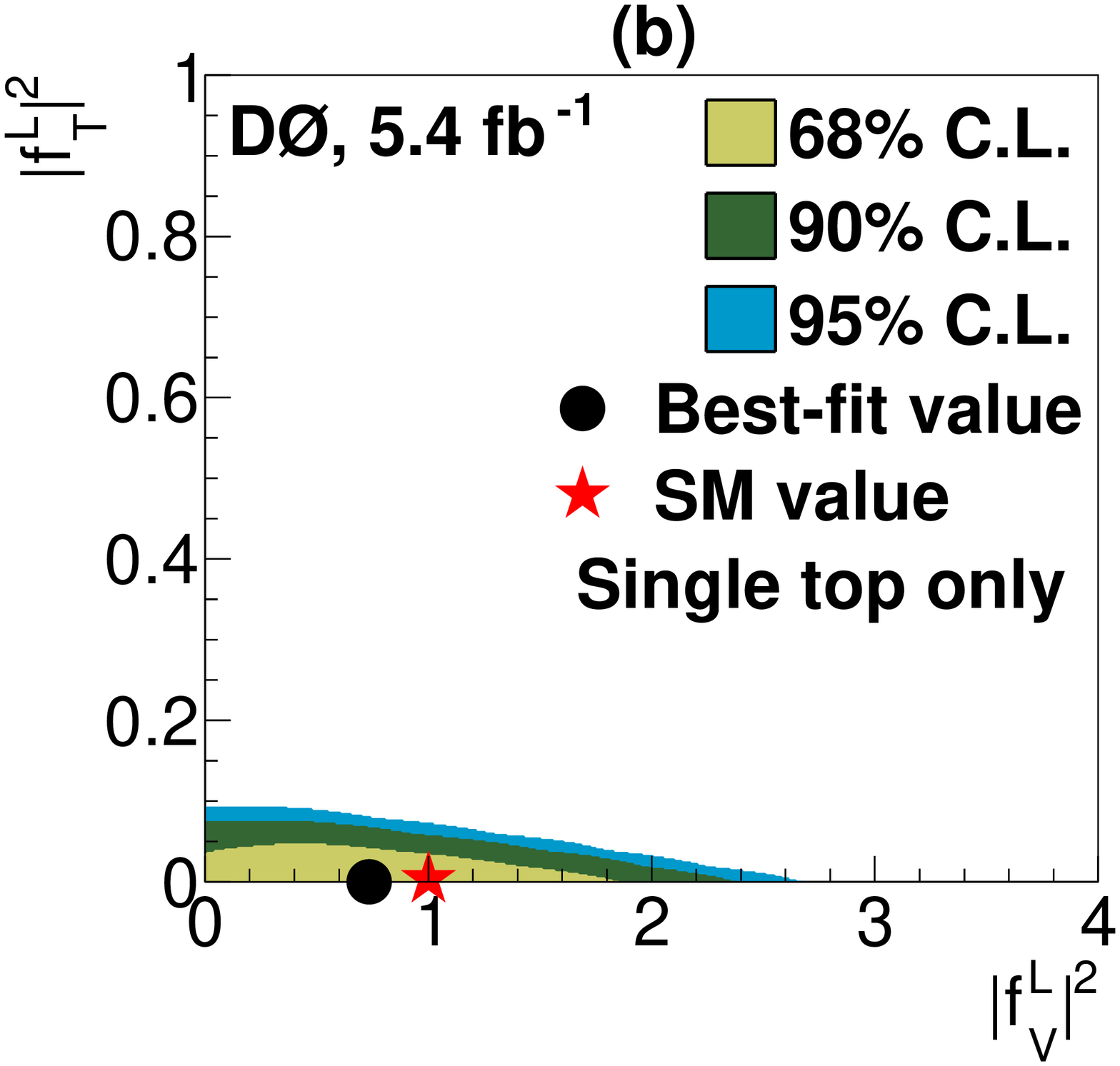}
\includegraphics[width=0.32\textwidth]{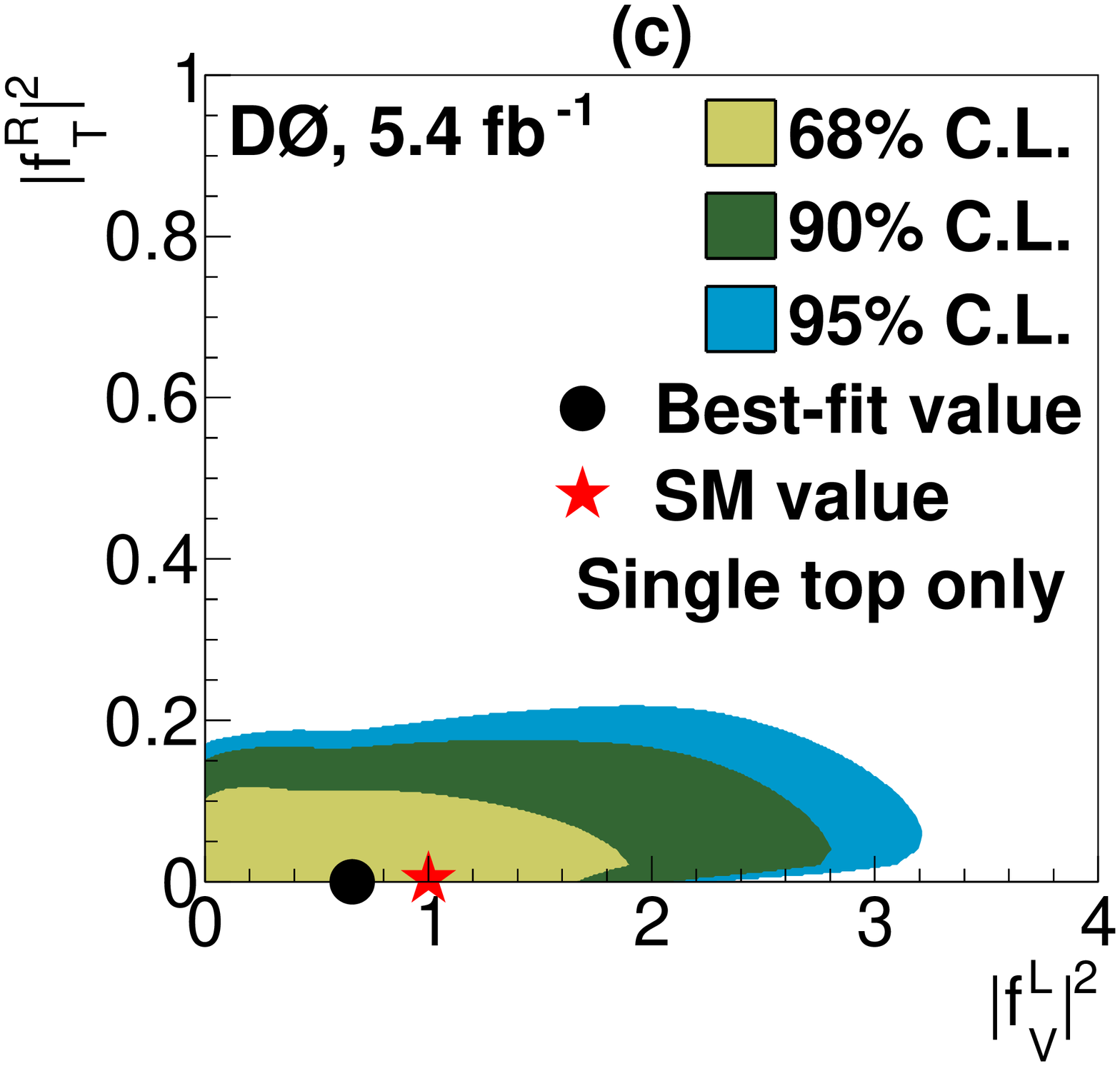}
\vspace{-0.1in}
\caption{(color online) Form factor posterior density distribution
for (a) right-vector vs. left-vector couplings,
(b) left-tensor vs. left-vector couplings and
(c) right-tensor vs. left-vector couplings, using information
from the single top quark analysis only, for events with two or three jets. 
All systematic uncertainties are included.}
\label{fig:posterior_noWhel}
\end{figure*}

The BNN output is used in a Bayesian analysis that determines a posterior density as 
a function of the anomalous coupling and the SM coupling, separately for each scenario.
Figure~\ref{fig:posterior_noWhel} shows the probability density distributions from the
single top quark anomalous couplings search, and the middle column of 
Table~\ref{tab:obslim} gives the anomalous coupling form factor 
limits obtained  from the single top quark anomalous
couplings analysis alone. These differ slightly from 
those given in Ref.~\cite{Abazov:2011pm} due to the exclusion of the 4-jet sample.

We account for all systematic uncertainties and their correlations among different
analysis channels, and sources of signal or background, in the two analyses.
Systematic uncertainties in the $W$~boson helicity measurement are detailed in
Ref.~\cite{Abazov:2010jn}. They arise from finite MC statistics and uncertainties on the 
top quark mass, jet energy scale, and MC models of 
signal and background.  Variations in these parameters can change the measurement in 
two ways: by altering the estimate of the background (i.e., if the background 
selection efficiency changes) and by modifying the shape
of the $\cos\theta^*$~templates. Systematic uncertainties on the $\ttbar$ normalization
do not affect the measurement.
We also assign a systematic uncertainty to account for differences between 
the input $f_0$ and $f_+$ values and the average fit values in pseudo-experiments.

Systematic uncertainties on the signal and background models in the single top quark
anomalous couplings analysis are
estimated using the methods described in Ref.~\cite{Abazov:2011pt}. The dominant sources
of uncertainty are the jet energy scale, $b$-tag modeling, and MC models of signal and 
background, with smaller contributions from background normalizations, top quark mass, and 
object identification. 

Uncertainties that only affect the $W$~boson helicity measurement are MC statistics and the
$\ttbar$ $\cos\theta^*$~template modeling uncertainty.
Uncertainties that only affect the single top quark anomalous 
coupling analysis are those related to signal modeling and background normalization,
including luminosity, object reconstruction, and $b$-tag modeling.

We use a Bayesian statistical analysis~\cite{bayes-limits} to combine the $W$~boson helicity 
result with that of the single top quark anomalous couplings analysis. The likelihood 
from the $W$~boson helicity analysis shown in Fig.~\ref{fig:whel_prior} is used as a prior to 
the analysis of single top anomalous couplings analysis.
For each anomalous coupling form factor scenario ($f^R_V$, $f^L_T$, and $f^R_T$), we compare
the corresponding BNN output for data with the sum of backgrounds and two signal models,
the anomalous coupling model and the SM ($f^L_V$). 
In the $f^L_T$ scenario the two amplitudes interfere for single top quark production, which is 
taken into account through a superposition of three signal samples: 
one with only left-handed vector couplings, one with only left-handed tensor 
couplings, and one with both coupling form factors set to one (which also
includes the interference term). For {\ttbar} production all interference terms are accounted 
for properly in all three scenarios.

We then compute a likelihood as a product 
over all separate analysis channels. We assume Poisson distributions for the observed 
counts and use Gaussian distributions to model the uncertainties on the 
signal acceptance and background yields, including correlations of systematic uncertainties.
The uncertainties are evaluated through MC integration in an ensemble of 200,000 samples.
Each sample has the same data distribution but signal and background contributions that
are shifted by the systematic uncertainties, 
i.e., the signal and background shapes and normalizations as well as the prior from the 
$W$~boson helicity change for
each sample. The final posterior is  the ensemble average of all individual posteriors. 

\begin{figure*}[!h!btp]
\includegraphics[width=0.32\textwidth] {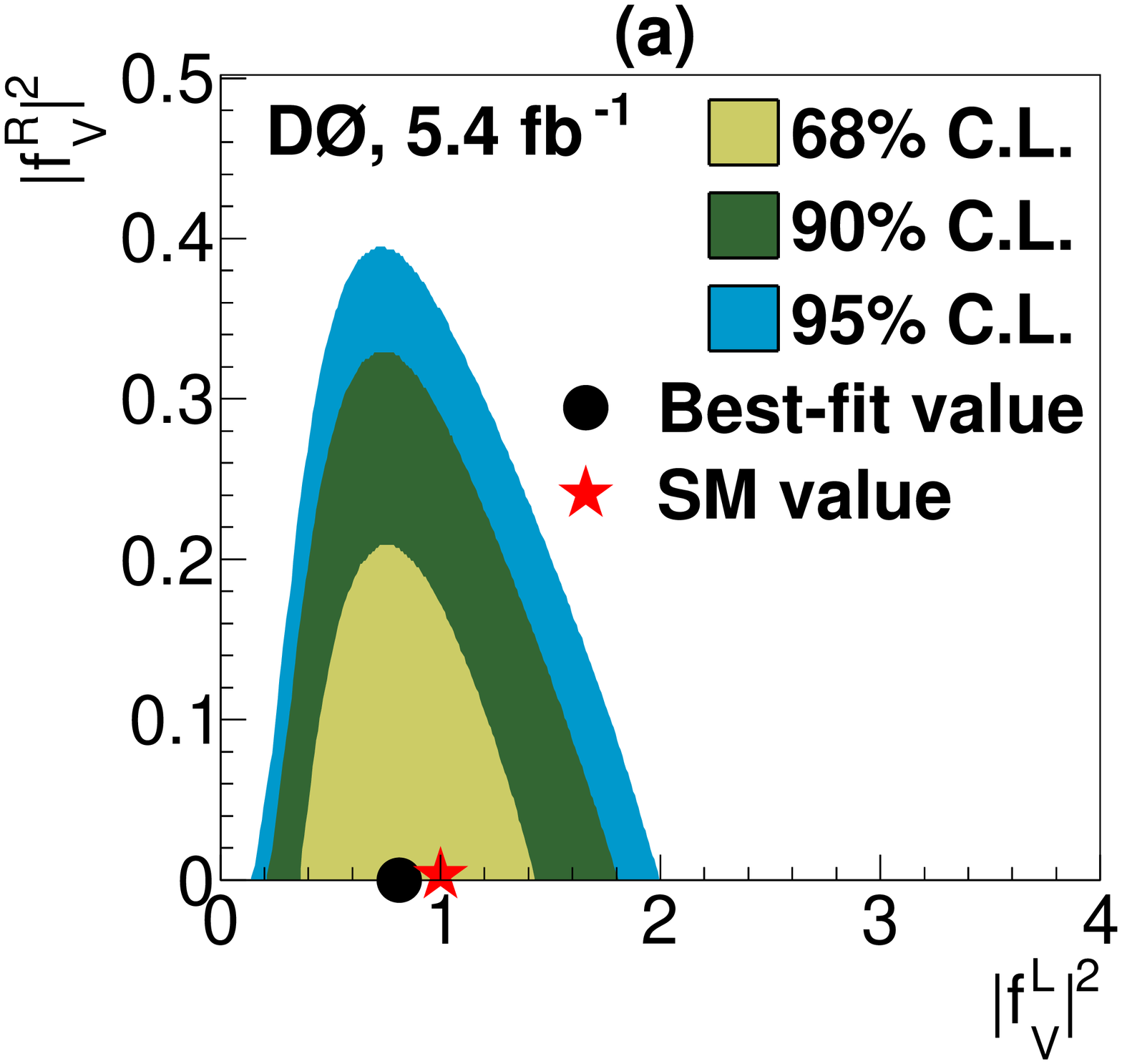}
\includegraphics[width=0.32\textwidth] {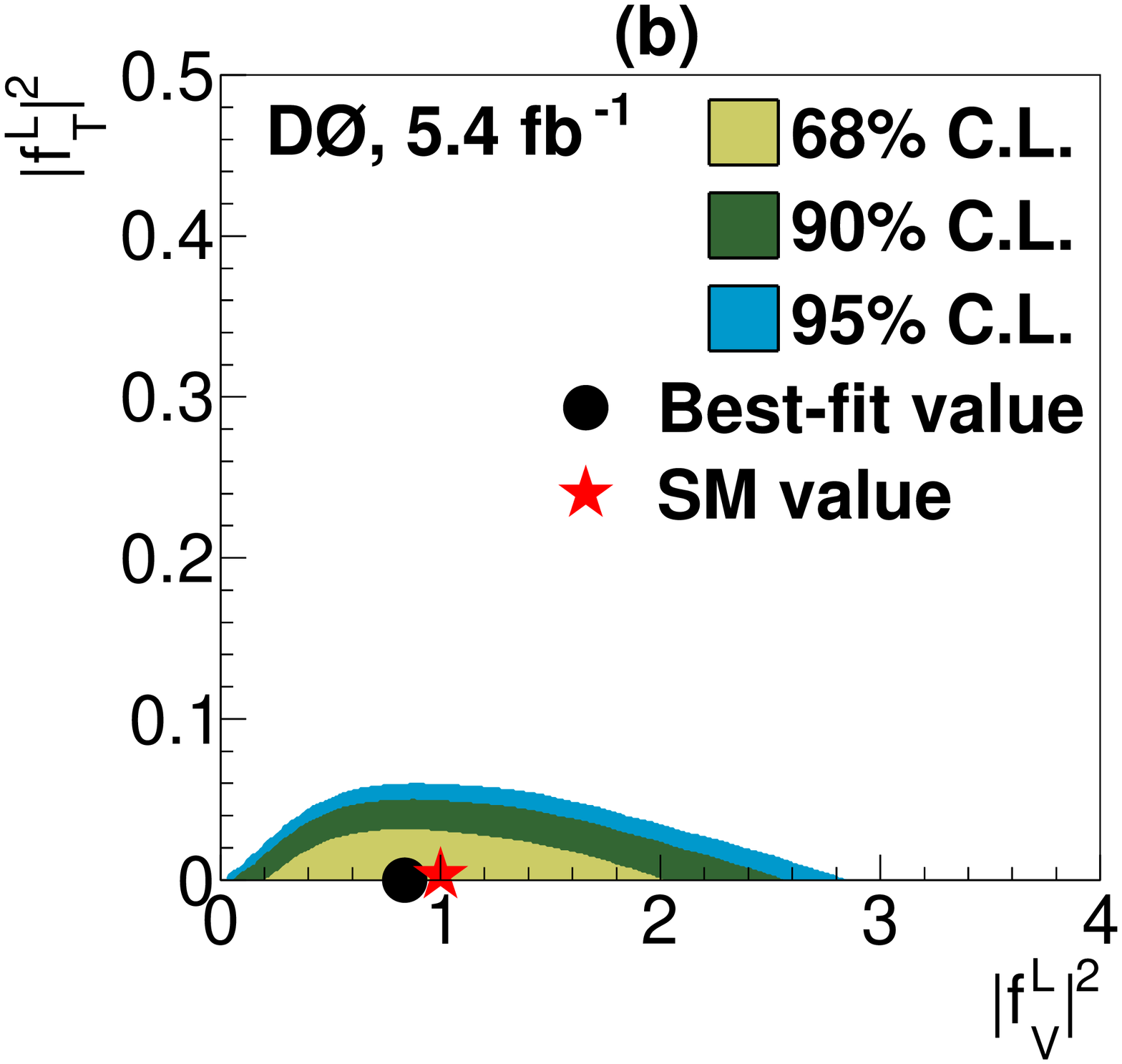}
\includegraphics[width=0.32\textwidth] {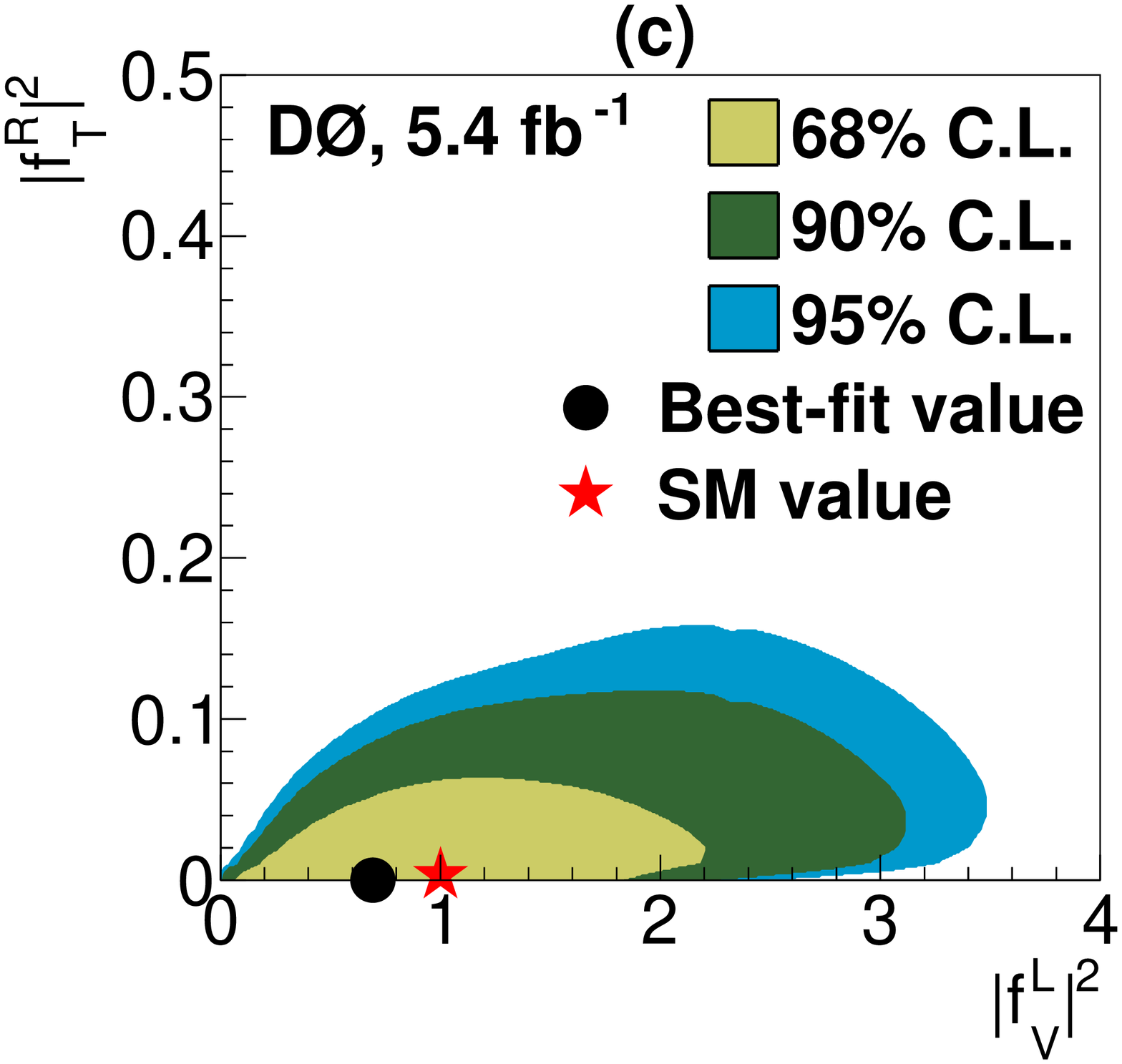}
\vspace{-0.1in}
\caption{Posterior density distribution for the combination of $W$~boson helicity
and single top quark measurements for (a) right-vector vs. left-vector form factors, 
(b) left-tensor vs. left-vector form factors and
(c) right-tensor vs. left-vector form factors. All systematic uncertainties are included.
}
\label{fig:measfullsys_2D}
\end{figure*}

The two-dimensional posterior probability density is computed as a function of 
$|f^L_V|^2$ and $|f_X|^2$, where $f_X$ is $f^R_V$, $f^L_T$, or $f^R_T$. 
These probability density distributions including both $W$~boson helicity and single top 
quark anomalous coupling information are shown in~Fig.~\ref{fig:measfullsys_2D}. 
We observe no significant anomalous contributions.

We compute 95\%~C.L. upper limits on the anomalous form factors by integrating over the 
left-handed vector contribution to obtain one-dimensional posterior probability densities. 
The limits are given in Table~\ref{tab:obslim}. 

\begin{table}[!h!t]
\caption{\label{tab:obslim} Observed upper limits on anomalous $tWb$ couplings
at 95\% C.L. from $W$~boson helicity assuming $f^L_V=1$, from the single top quark analysis, 
and from their combination, for which no assumption on $f^L_V$ is made.
}
\begin{tabular}{c|c|c|c}
\hline \hline
Scenario  & only             &  only              & combination \\
          & $W$~helicity     &  single top        & \\
\hline	
$|f_V^R|^2$ & $ 0.62 $ & $ 0.89 $ & $ 0.30 $ \\
$|f_T^L|^2$ & $ 0.14 $ & $ 0.07 $ & $ 0.05 $ \\
$|f_T^R|^2$ & $ 0.18 $ & $ 0.18 $ & $ 0.12 $ \\
\hline \hline
\end{tabular}
\end{table}

Table~\ref{tab:obslim} also shows the limits obtained from only the $W$~boson helicity
analysis with the additional assumption that $f^L_V=1$. 
Compared with the results obtained using only the single-top search,
the combination improves the limits on the form factors significantly because
the individual analyses provide complementary information.

The 95\% C.L. limits on the coupling operators in the operator notation 
based on Eq.~\ref{eq:operator} are
$|C_{\phi q}^{(3,3+3)}| < 14.7$, $|C_{\phi \phi}^{33}|< 18.0$,
$|C_{dW}^{33}|< 2.5$, and $|C_{uW}^{33}|< 4.1$,
assuming a new physics scale of $\Lambda=1$~TeV. The limit on $C_{\phi q}^{(3,3+3)}$ is 
obtained from the $f^R_V$ scenario filter by setting $f^R_V=0$
and integrating the resulting $|f^L_V|^2$ posterior density starting at $|f^L_V|^2=1$
to find the 95\% C.L. limit on the anomalous contribution.
Limits for the other operators are obtained from the corresponding form factor limits.
These limits are a significant improvement over previous limits. 
A separate analysis of Tevatron and early LHC results~\cite{AguilarSaavedra:2011ct} provides
limits on anomalous couplings that appear stronger than those presented here even though it
uses less information. This is mainly due to the use of priors that are flat 
in the coupling rather than the coupling squared as is done here.

%
In summary, we have presented a study of $tWb$ couplings that combines $W$~boson helicity 
measurements in top quark decay with anomalous couplings searches in the single top quark 
final state, thus using all currently applicable top quark measurements by D0. We find 
consistency 
with the SM and set 95\%~C.L. limits on anomalous $tWb$ couplings. Our limits represent 
significant improvements over previous D0 results beyond the increase in luminosity.

%
We thank the staffs at Fermilab and collaborating institutions,
and acknowledge support from the
DOE and NSF (USA);
CEA and CNRS/IN2P3 (France);
MON, Rosatom and RFBR (Russia);
CNPq, FAPERJ, FAPESP and FUNDUNESP (Brazil);
DAE and DST (India);
Colciencias (Colombia);
CONACyT (Mexico);
NRF (Korea);
FOM (The Netherlands);
STFC and the Royal Society (United Kingdom);
MSMT and GACR (Czech Republic);
BMBF and DFG (Germany);
SFI (Ireland);
The Swedish Research Council (Sweden);
and
CAS and CNSF (China).
%


\end{document}